\documentclass[twocolumn,showpacs,aps,prl,superscriptaddress]{revtex4}

\usepackage{xspace}
\usepackage{graphicx}
\usepackage{dcolumn}
\usepackage{amsmath}
\usepackage{epsfig}

\input babarsym.tex

\newcommand{\lumi}    {91.5\invfb}
\newcommand{\lumion}  {81.9\invfb}
\newcommand{\lumioff} {9.6\invfb}

\newcommand{\lll}     {\ensuremath{\ell^{-}\ell^{+}\ell^{-}}}
\newcommand{\eee}     {\ensuremath{e^-\!e^+\!e^-}}
\newcommand{\eemw}    {\ensuremath{\mu^+\!e^-\!e^-}}
\newcommand{\eemr}    {\ensuremath{\mu^-\!e^+\!e^-}}
\newcommand{\emmw}    {\ensuremath{e^+\!\mu^-\!\mu^-}}
\newcommand{\emmr}    {\ensuremath{e^-\!\mu^+\!\mu^-}}
\newcommand{\mmm}     {\ensuremath{\mu^-\!\mu^+\!\mu^-}}

\newcommand{\taulll}  {\ensuremath{\tau^{-}\!\to\lll}}
\newcommand{\dEdM}    {\ensuremath{(\Delta M, \Delta E)}}

\def\kk2f       {\mbox{\tt KK2f}\xspace}
\def\tauola     {\mbox{\tt Tauola}\xspace}
\def\photos     {\mbox{\tt Photos}\xspace}
\newcommand{\Nobs}      {\ensuremath{N_{\rm obs}}}
\newcommand{\Nbgd}      {\ensuremath{N_{\rm bgd}}}
\newcommand{\Nul}       {\ensuremath{N_{\rm UL}^{90}}}
\newcommand{\BRul}       {\ensuremath{\BR_{\rm UL}^{90}}}
\newcommand{\tensev}    {\ensuremath{\times 10^{-7}}}

\newcommand{\BABARPubYear}     {03}
\newcommand{\BABARPubNumber}  {044}

\newcommand{\SLACPubNumber} {10256}

\def\figurebox#1#2#3{%
    \def\arg{#3}%
    \ifx\arg\empty
    {\hfill\vbox{\hsize#2\hrule\hbox to #2{\vrule\hfill\vbox to #1{\hsize#2\vfill}\vrule}\hrule}\hfill}%
    \else
    {\hfill\epsfbox{#3}\hfill}%
    \fi}

\begin{document}

\preprint{\babar-PUB-\BABARPubYear/\BABARPubNumber} 
\preprint{SLAC-PUB-\SLACPubNumber} 

\begin{flushleft}
\babar-PUB-\BABARPubYear/\BABARPubNumber\\
SLAC-PUB-\SLACPubNumber\\
\end{flushleft}

\title{
{\large \bf \boldmath
Search for Lepton-Flavor Violation in the Decay \taulll} 
}

%
\author{B.~Aubert}
\author{R.~Barate}
\author{D.~Boutigny}
\author{F.~Couderc}
\author{J.-M.~Gaillard}
\author{A.~Hicheur}
\author{Y.~Karyotakis}
\author{J.~P.~Lees}
\author{V.~Tisserand}
\author{A.~Zghiche}
\affiliation{Laboratoire de Physique des Particules, F-74941 Annecy-le-Vieux, France }
% ...
\collaboration{The \babar\ Collaboration}
\noaffiliation

\date{\today}

\begin{abstract}
A search for the lepton-flavor-violating decay of the 
tau into three charged leptons has been performed 
using \lumi\ of data collected at an \epem
center-of-mass energy of 10.58\gev with the \babar\ detector 
at the \pep2\ storage ring.
In all six decay modes considered, the numbers of events found 
in data are compatible with the background expectations.
Upper limits on the branching fractions are set in the range 
$(1-3) \times10^{-7}$ at 90\% confidence level.
\end{abstract}

\pacs{13.35.Dx, 14.60.Fg, 11.30.Hv}

\maketitle


Lepton-flavor violation (LFV) involving charged leptons has 
never been observed, and stringent experimental limits 
exist from muon branching fractions:
$\BR(\mmu\to\electron\gamma) < 1.2 \times 10^{-11}$ \cite{brooks99}
and $\BR(\mmu\to\electron\electron\electron) < 1.0 
\times 10^{-12}$ \cite{sindrum88} at 90\% confidence level (CL).
Recent results from neutrino oscillation experiments \cite{neut} 
show that LFV does indeed occur, although the branching fractions 
expected in charged lepton decay due to neutrino mixing alone 
are probably no more than $10^{-14}$ \cite{pham98}.

In tau decays, the most stringent limit on LFV is 
$\BR(\mtau\to\mmu\gamma) < 3.1 \times 10^{-7}$ 
at 90\% CL \cite{belle03}.
Many extensions to the Standard Model (SM), particularly models 
seeking to describe neutrino mixing, predict enhanced LFV in tau 
decays over muon decays with branching fractions from 
$10^{-10}$ up to the current experimental limits \cite{ma02}.
Observation of LFV in tau decays would be a 
clear signature of non-SM physics, while improved 
limits will provide further constraints on theoretical models.

This analysis is based on data recorded 
by the \babar\ detector at the \pep2\ asymmetric-energy \epem\ 
storage ring operated at the Stanford Linear Accelerator Center.
The data sample consists of \lumion\ recorded at
$\sqrt{s} = 10.58 \gev$ and \lumioff\ recorded at
$\sqrt{s} = 10.54 \gev$.
With an expected cross section for tau pairs at the luminosity-weighted 
$\sqrt{s}$ of $\sigma_{\tau\tau} = (0.89\pm0.02)$ nb \cite{kk},
this data sample contains over 160 million tau decays.
 
The \babar\ detector is described in detail in Ref.~\cite{detector}.
Charged-particle (track) momenta are measured with a 5-layer
double-sided silicon vertex tracker and a 40-layer drift chamber 
inside a 1.5-T superconducting solenoidal magnet.
The transverse momentum resolution parameterized as
$\sigma_{\pt}/\pt = (0.13\cdot \pt/[\gevc] + 0.45)\%$ is achieved.
An electromagnetic calorimeter consisting of 6580 CsI(Tl) 
crystals is used to identify electrons and photons,
a ring-imaging Cherenkov detector is used to identify
charged hadrons, 
and the instrumented magnetic flux return (IFR) is used to
identify muons.
Particle attributes are reconstructed in the laboratory
frame and then boosted to the \epem\ center-of-mass (CM) 
frame using the measured asymmetric beam energies.

This paper presents a search for LFV in the neutrinoless
decay \taulll.
All possible lepton combinations consistent with charge
conservation are considered, leading to six distinct
decay modes (\eee, \eemw, \eemr, \emmw, \emmr, \mmm)
\cite{cc}.
The signature of this process is three charged 
particles, each identified as either an electron or muon,
with an invariant mass and energy equal to that of the parent 
tau lepton.
Candidate signal events in this analysis are required
to have a ``1-3 topology,'' where one tau decay yields three
charged particles (3-prong), while the second tau
decay yields one charged particle (1-prong).
Four well reconstructed tracks are required 
with zero net charge, 
pointing towards a common region consistent with 
\tautau production and decay.
One of these tracks must be separated from the other 
three by at least 90\degrees in the CM frame.
The plane perpendicular to this isolated track divides 
the event into two hemispheres and defines the 1-3 topology.
Pairs of oppositely charged tracks identified as
photon conversions in the detector material with 
an \epem invariant 
mass below 30\mevcc are ignored.

Each of the charged particles found in the 3-prong 
hemisphere must be identified as either an electron
or muon candidate.
Electrons are identified using the ratio of
calorimeter energy to track momentum $(E/p)$, the ionization 
loss in the tracking system $(\dedx)$, and the shape of the shower
in the calorimeter.
Muons are identified by hits in the IFR
and small energy deposits in the
calorimeter.
Muons with momentum less than $0.5\gevc$ cannot be identified
because they do not penetrate far enough into the IFR.

The particle identification (PID) requirements
are not sufficient to suppress certain backgrounds, 
particularly those from higher-order radiative 
Bhabha and \mumu events that can have four leptons 
in the final state.
To reduce these backgrounds, additional selection criteria are
applied to the six different decay modes.  
For all decay modes, the momentum of the 1-prong track is required 
to be less than 4.8 GeV/c in the CM frame.
For the \eee\ and \emmr\ decay modes, the charged particle
in the 1-prong hemisphere must not be identified as an
electron, while for the \eemr\ and \mmm\ decay modes
it must not be a muon.
For all four of these decay modes, the angle $\theta_{13}$
between the 1-prong
momentum and the vector sum of the 3-prong momenta in
the CM frame must satisfy $\cos\theta_{13}>-0.9999$,
while the net transverse momentum of the four tracks
must be greater than 0.1\gevc.
Additional requirements are imposed to reduce the \qqbar
and SM \tautau backgrounds.
Events in the four decay modes specified above are
required to have no unassociated calorimeter clusters
(photons) in the 3-prong hemisphere with energy 
greater than $100\mev$ in the laboratory frame, while 
events in all six decay modes are required to have 
no track in the 3-prong hemisphere that is also
consistent with being a kaon.

To reduce backgrounds further,
signal events are required to have
an invariant mass and total energy in the 3-prong
hemisphere consistent with a parent tau lepton.
These quantities are calculated from the observed track momenta 
assuming the corresponding lepton masses for each decay mode.
The energy difference is defined as 
$\Delta E \equiv E^{\star}_{\mathrm{rec}} - E^{\star}_{\mathrm{beam}}$,
where $E^{\star}_{\mathrm{rec}}$ is the total energy of the tracks
observed in the 3-prong hemisphere and $E^{\star}_{\mathrm{beam}}$
is the beam energy, both in the CM frame.
The mass difference is defined as
$\Delta M \equiv M_{\mathrm{rec}} - m_{\tau}$ where $M_{\mathrm{rec}}$ 
is the reconstructed invariant mass of the three tracks
and $m_{\tau}=1.777\gevcc$ is the tau mass \cite{bes}.

\begin{figure}
 \resizebox{\columnwidth}{!}{%
\includegraphics{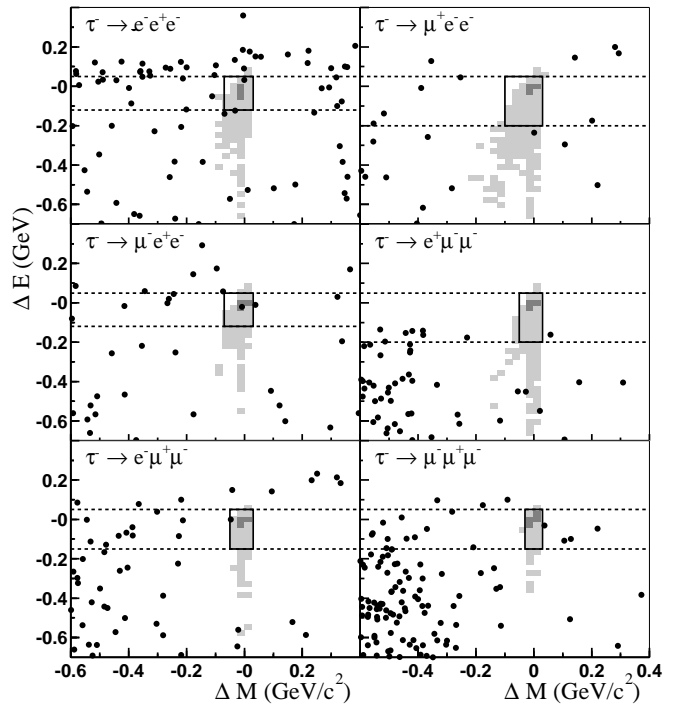}}
\caption{Observed data shown as dots in the \dEdM\ plane and 
the boundaries of the signal region for each decay mode.
The dark and light shading indicates contours containing
50\% and 90\% of the selected MC signal events, respectively.
The regions shown in Fig.~\ref{fig2} are indicated by dashed lines.}
\label{fig1}
\end{figure}

The signal distributions in the \dEdM\ plane are 
broadened by detector resolution and radiative effects.
The radiation of photons from the incoming \epem\ particles
before annihilation affects all decay modes, leading to a tail
at low values of $\Delta E$.
Radiation from the final-state leptons 
is more likely for electrons than muons, and produces a tail
at low values of $\Delta M$ as well.
Rectangular signal regions are defined separately for each 
decay mode as follows.
For all six decay modes, the upper right corner of
the signal region is fixed at $(30\mevcc, 50\mev)$,
while the lower left corner is at $(-70, -120)$ 
for the \eee\ and \eemr\ decay modes, $(-100, -200)$
for \eemw, $(-50, -200)$ for \emmw, $(-50, -150)$
for \emmr, and $(-30, -150)$ for \mmm.
All values are given in units of $(\mevcc, \mev)$.
These signal region boundaries are chosen
to provide the smallest expected upper 
limits on the branching fractions in the background-only hypothesis.
These expected upper limits are estimated using only Monte Carlo (MC)
simulations and data control samples, not candidate signal events.
Figure~\ref{fig1} shows the observed data in the \dEdM\ plane, 
along with the signal region boundaries 
and the expected signal distributions.
To avoid bias, a blinded analysis procedure was adopted
with the number of data events in the signal region
remaining unknown until the selection criteria 
were finalized and all cross checks were performed.

The efficiency of the selection for signal events is
estimated with a MC simulation of LFV tau decays.
Simulated tau-pair events including higher-order radiative
corrections are generated using \kk2f \cite{kk}
with one tau decaying to three leptons with a 3-body 
phase space distribution, while the other tau decays 
according to measured rates \cite{PDG} simulated with \tauola \cite{tauola}.
Final state radiative effects are simulated for all decays 
using \photos \cite{photos}.
The detector response is simulated with \mbox{\tt GEANT4}~\cite{geant},
and the simulated events are then reconstructed in the same 
manner as data.

About 50\% of the MC signal events pass the 1-3 topology requirement.
The lepton identification efficiencies and misidentification 
probabilities are measured using tracks in kinematically-selected
data samples
(radiative Bhabha, radiative \mumu, two-photon $\epem\ellell$, and $\jpsi\to\ellell$) and parameterized as a 
function of particle momentum, polar angle, and azimuthal angle
in the laboratory frame.
These data-derived efficiencies are then used to give the
probability that a simulated MC particle will be identified 
(or misidentified) as an electron or a muon.
For the lepton momentum spectrum predicted by the signal MC,
the electron and muon identification requirements are found to have an
average efficiency per lepton of 91\% and 63\%, respectively.
The probability for a hadron to be misidentified as an electron in 
SM 3-prong tau decays is 2.2\%, while the probability to be misidentified
as a muon is 4.8\% \cite{pid}.
The final efficiency for signal events to be found
in the signal region is shown in Table~\ref{tab:results} for
each decay mode and ranges from 7\% to 12\%.
This efficiency includes the 85\% branching fraction for 1-prong 
tau decays.

\begin{table}
\begin{center}
\caption{Efficiency estimates, number of expected background events (\Nbgd),
number of observed events (\Nobs), and branching fraction upper limits 
for each decay mode.  
}
\begin{tabular}{lccc}
\hline\hline
Decay mode   & \eee         & \eemw        & \eemr \\
\hline
Efficiency [\%] &$7.3\pm 0.2  $&$11.6\pm 0.4 $&$7.7\pm 0.3$ \\
\hline
\qqbar bgd.     &$0.67        $&$0.17        $&$0.39$ \\
QED bgd.        &$0.84        $&$0.20        $&$0.23$ \\
\tautau bgd.    &$0.00        $&$0.01        $&$0.00$ \\
\hline
\Nbgd           &$1.51\pm 0.11$&$0.37\pm 0.08$&$0.62\pm 0.10$ \\
\Nobs           &$1           $&$0           $&$1$ \\
\hline
\BRul            &$2.0\tensev  $&$1.1\tensev  $&$2.7\tensev$ \\
\hline
\hline
Decay mode      & \emmw        & \emmr        & \mmm \\
\hline
Efficiency [\%] &$9.8\pm 0.5  $&$6.8\pm 0.4  $&$6.7\pm 0.5$ \\
\hline
\qqbar bgd.     &$0.20        $&$0.19        $&$0.29$ \\
QED bgd.        &$0.00        $&$0.19        $&$0.01$ \\
\tautau bgd.    &$0.01        $&$0.01        $&$0.01$ \\
\hline
\Nbgd           &$0.21\pm 0.07$&$0.39\pm 0.08$&$0.31\pm 0.09$ \\
\Nobs           &$0           $&$1           $&$0$ \\
\hline
\BRul            &$1.3\tensev  $&$3.3\tensev  $&$1.9\tensev$ \\
\hline
\hline
\end{tabular}
\label{tab:results}
\end{center}
\end{table}

There are three main classes of background remaining after
the selection criteria are applied: low multiplicity \qqbar events
(mainly continuum light-quark production), QED events (Bhabha 
and \mumu), and SM \tautau events.
These three background classes have distinctive distributions
in the \dEdM\ plane:
\qqbar events tend to populate the plane uniformly,
while QED backgrounds are restricted to a narrow band
at positive values of $\Delta E$, and \tautau backgrounds
are restricted to negative values of both $\Delta E$ and $\Delta M$.
A negligible two-photon background remains.

The expected background rates for each decay mode are determined by
fitting a set of probability density functions (PDFs) to the
observed data in the \dEdM\ plane in a grand sideband (GS) region.
The GS region, shown in Fig.~\ref{fig1},
is defined as the rectangle bounded by the points
$(-600\mevcc, -700\mev)$ and $(400\mevcc, 400\mev)$, excluding
the signal region.
For both the \qqbar and \tautau backgrounds, an analytic PDF is
constructed from the product of two PDFs $P_M$ and 
$P_E$, where
$P_M(\Delta M)$ is the sum of two Gaussians with a common mean and 
$P_E(\Delta E) = (1-x/\sqrt{1+x^2})(1+a x+b x^2+c x^3)$ with $x=(\Delta E-d)/e$
\cite{opal00}.
The shapes of these PDFs are described by a total of nine free 
parameters, which are determined by fits to MC \qqbar and \tautau
background samples for each decay mode.

For the QED backgrounds, an analytic PDF is constructed from
the product of a Crystal Ball function \cite{CBF} in $\Delta E'$ 
and a linear function in $\Delta M'$, where the
$(\Delta M', \Delta E')$ axes have been rotated slightly
from \dEdM\ to fit the observed distribution.
The six parameters of this PDF, including the rotation angle,
are obtained by fitting control samples with a 1-3 topology
that are enhanced in Bhabha or \mumu events by requiring that 
the particle in the 1-prong hemisphere is identified as an electron 
or muon.
Any value for $\cos\theta_{13}$ is allowed, but the control sample 
events otherwise pass the selection criteria.

\begin{figure}
 \resizebox{\columnwidth}{!}{%
\includegraphics{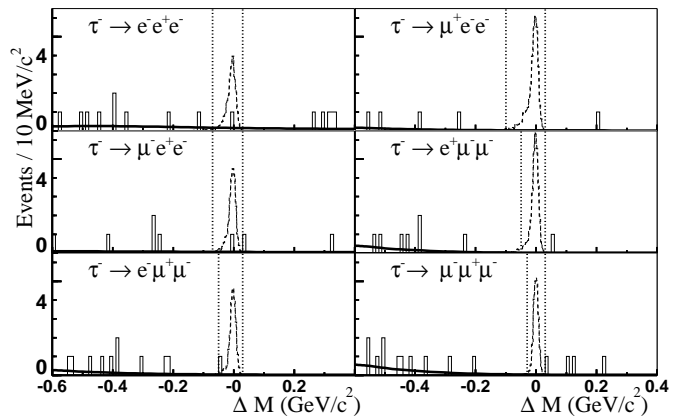}}
\caption{Distribution of $\Delta M$ for data (solid histogram) and
background PDFs (solid curves) for events with $\Delta E$ in the
signal region indicated in Fig.~\ref{fig1}.
Expected signal distributions are shown (dashed histogram)
for a branching fraction of \mbox{$10^{-6}$}.}
\label{fig2}
\end{figure}

With the shapes of the three background PDFs determined, 
an unbinned maximum likelihood fit to the data in the GS region
is used to find the expected rate of each background type
in the signal region, as shown in Table~\ref{tab:results}.
The PDF shape determinations and background fits are performed
separately for each of the six decay modes.
Figure~\ref{fig2} shows the data and the background PDFs
for values of $\Delta E$ in the signal range.

The largest systematic uncertainty in 
the signal efficiency is due to the uncertainty in
measuring the PID efficiencies.
This uncertainty is determined from the statistical precision
of the PID control samples, 
and ranges from $0.7\%$ for \eee\ to $6.2\%$ for \mmm relative
to the efficiency \cite{uncertain}.
The modeling of the tracking efficiency contributes an additional 2\%
uncertainty, as does the statistical limitation of the MC signal sample.
All other sources of uncertainty are found to be 
small, including the modeling in the generator of radiative effects, 
track momentum resolution, trigger performance, observables used in 
the selection criteria, and knowledge of the tau 1-prong branching 
fractions.
The efficiency has been estimated using a 3-body phase space model
and no uncertainty is assigned for possible model dependence.
The selection efficiency is found to be uniform within 
10\% across the Dalitz plane, provided
the invariant mass for any pair of leptons is 
less than $1.4 \gevcc$.

Since the background levels are extracted directly from the data,
systematic uncertainties on the background estimation are directly
related to the background parameterization and the fit technique used.
The finite data available in the GS region to determine the
background rates is the largest uncertainty and varies
from 10\% to 25\% depending upon the decay mode.
Additional uncertainties are estimated by varying the fit 
procedure and changing the functional form of the background PDFs.
Cross checks of the background estimation were performed by 
considering the number of events expected and observed in 
sideband regions immediately neighboring the signal region for 
each decay mode.

The numbers of events observed (\Nobs) and the 
background expectations (\Nbgd) are shown in Table~\ref{tab:results}, 
with no significant excess found in any decay mode.
Upper limits 
on the branching fractions are calculated according to 
$\BRul = \Nul/(2 \varepsilon \L \sigma_{\tau\tau})$, where $\Nul$
is the 90\% CL upper limit for the number 
of signal events when \Nobs\ events are observed with \Nbgd\ background 
events expected.
The values $\varepsilon$, $\L$, and $\sigma_{\tau\tau}$ are the
selection efficiency, luminosity, and \tautau cross section, respectively.
The estimates of $\L = \lumi$ and $\sigma_{\tau\tau} = 0.89$ nb are 
correlated \cite{lum}, and the uncertainty on the product 
$\L \sigma_{\tau\tau}$ is 2.3\%.
The branching fraction upper limits have been calculated including
all uncertainties using the technique of 
Cousins and Highland \cite{cousins92} following the implementation of 
Barlow \cite{barlow02}.
The 90\% CL upper limits on the \taulll\ branching fractions, shown
in Table~\ref{tab:results}, are in the range $(1-3)\times10^{-7}$.

We are grateful for the excellent luminosity and machine conditions
provided by our \pep2\ colleagues, 
and for the substantial dedicated effort from
the computing organizations that support \babar.
The collaborating institutions wish to thank 
SLAC for its support and kind hospitality. 
This work is supported by
DOE
and NSF (USA),
NSERC (Canada),
IHEP (China),
CEA and
CNRS-IN2P3
(France),
BMBF and DFG
(Germany),
INFN (Italy),
FOM (The Netherlands),
NFR (Norway),
MIST (Russia), and
PPARC (United Kingdom). 
Individuals have received support from the 
A.~P.~Sloan Foundation, 
Research Corporation,
and Alexander von Humboldt Foundation.


\begin{thebibliography}{99}
\bibitem{cc}
Throughout this paper, charge conjugate decay modes also are implied.
\bibitem{uncertain}
All uncertainties quoted in the text are relative.
\bibitem{lum}
The luminosity is measured using the observed \mumu rate, and
the \mumu and \tautau cross sections are both estimated with {\tt KK2f}.
\end{thebibliography}
\end{document}